\documentclass[prl,twocolumn,superscriptaddress,footinbib]{revtex4-1}
\usepackage{amsmath,amssymb,bm}
\usepackage{graphicx}
\usepackage{epstopdf}
\usepackage{latexsym}
\usepackage{subfigure}
\usepackage{color}

\begin{document}

\title{A lattice model for the SU($N$) N\'eel-VBS quantum phase transition at large $N$}

\author{Ribhu K. Kaul} 
\affiliation{Department of Physics \& Astronomy, University of Kentucky, Lexington, KY-40506-0055}

\author{Anders W. Sandvik}
\affiliation{Department of Physics, Boston University, 590 Commonwealth Avenue Boston, MA-02215}

\begin{abstract}
We generalize the SU($N=2$) $S=1/2$ square-lattice quantum magnet with nearest-neighbor antiferromagnetic coupling ($J_1$) 
and next-nearest-neighbor ferromagnetic coupling ($J_2$) to arbitrary $N$. For all $N>4$, the ground state has valence-bond-solid 
(VBS) order for $J_2=0$ and N\'eel order for $J_2/J_1\gg 1$, allowing
us access to the transition between these types of states 
for large $N$. Using quantum Monte Carlo simulations, we show that both order parameters vanish at a single quantum-critical 
point, whose universal exponents for large enough $N$ (here up to $N=12$) approach the values obtained in a $1/N$ expansion 
of the non-compact CP$^{N-1}$ field theory. These results lend strong support to the deconfined quantum-criticality theory
of the N\'eel--VBS transition.
\end{abstract}
\maketitle
 
Just as the destruction of magnetic order by thermal fluctuations is the paradigmatic 
example of a classical critical point \cite{chaikin2000:cmp}, the destruction of magnetic 
order at $T=0$ by quantum fluctuations is a prototypical example of a quantum-critical 
point \cite{sachdev1999:qpt}. Sometimes the quantum case can be entirely different due to novel 
quantum interference effects, which have no natural classical analogues. An important 
example is that of the square-lattice SU($N$) quantum antiferromagnet, where the destruction 
of SU($N$) symmetry-breaking N\'eel order in the background of uncompensated Berry phases
results in a valence-bond-solid (VBS) state with broken translational symmetry
\cite{haldane1988:berry,read1989:vbs,marston1988:sun}. Recent work has presented speculative 
albeit compelling arguments that a direct generically continuous N\'eel--VBS transition can exist. At such a  {\em deconfined} quantum critical (DQC) point both order parameters are simultaneously critical 
\cite{senthil2004:science}, a striking feature which is not contained in the conventional field-theory 
description of two independent order parameters (where one would instead
expect a generic direct transition to be first-order). 

Given the major paradigm shift that could be spawned by the DQC idea, it has been of great interest to verify 
its validity by unbiased numerical studies of lattice spin models (Hamiltonians) that harbor N\'eel--VBS 
transitions \cite{sandvik2007:deconf,melko2008:jq,jiang2008:first,lou2009:sun,kaul2011:su34}. 
The weight of evidence from such work indicates that generically continuous N\'eel--VBS transitions indeed 
exist in quantum spin systems, with initial skepticism \cite{jiang2008:first,kuklov2009:dejavu} appearing 
increasingly unfounded \cite{sandvik2010:logs,banerjee2010:log}.

This DQC scenario predicts that the SU($N$) N\'eel--VBS quantum-critical point falls into the universality class 
of the $(2+1)$-dimensional non-compact CP$^{N-1}$ field theory
\cite{motrunich2004:hhog,kamal1993:o3}. The 
connection between the phase transition in a microscopic Hamiltonian
and the low-energy continuum theory description relies on speculative assumptions
that are yet to be 
demonstrated convincingly. In order to provide support for this connection, therefore, one has to compare universal properties arising from these two starting points. 
Currently, the only technique with which the properties following from the CP$^{N-1}$ action can be studied analytically 
is the ${1}/{N}$ expansion \cite{halperin1974:largeN}, but it is not clear whether the results of this approach are 
valid down to the most interesting case of $N=2$. To test the DQC theory in an unbiased manner, it is thus of utmost 
importance to find quantum models in which the SU($N$) N\'eel-VBS transition can be studied for arbitrary large $N$, 
to compare the critical exponents with those of the $1/N$ expansion of the continuum field theory.

\begin{figure}
\includegraphics[width=6.5cm, trim = 20 155 0 0]{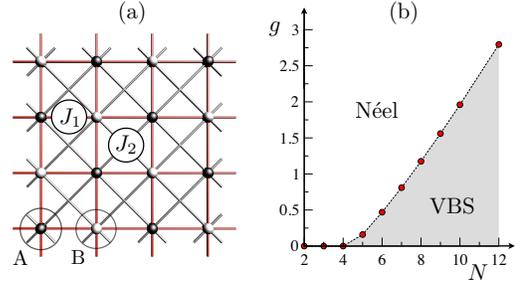}
\caption{\label{fig:intro} (Color online) (a) Black (white) lattice sites indicate the 
  A (B) sublattice on which spins of the $J_1$-$J_2$ model (\ref{eq:j1j2N}) transform as 
  the fundamental (conjugate) representation of SU($N$). $J_1$ couples nearest neighbors 
  with an SU($N$) {\em singlet projection} and  $J_2$ connects next nearest neighbors 
  with an SU($N$) {\em permutation}. (b) Phase diagram of the model as a function of 
  $N$ and $g\equiv J_2/J_1$.}
\vskip-3mm
\end{figure}

Until now, the N\'eel-VBS transition could be accessed in an unbiased manner only for $N\leq 4$, by quantum Monte 
Carlo (QMC) simulations of the so-called J-Q model 
\cite{sandvik2007:deconf,melko2008:jq,jiang2008:first,lou2009:sun,kaul2011:su34,banerjee2010:log}, in
which the $S=1/2$ Heisenberg ($J$) model is supplemented by certain multi-spin interactions ($Q$) favoring 
a VBS state. Is clear that this model alone cannot access the quantum transition for larger-$N$, however, simply 
because for $N\geq 5$ the $J$ model is itself VBS ordered \cite{harada2003:sun,beach2009:sun} and the $Q$ term only
increases the strength of this order. To remedy this problem, we here introduce an SU($N$) symmetric generalization
of the $J_1$-$J_2$ Heisenberg model model, with antiferromagnetic nearest-neighbor coupling $J_1$ and
ferromagnetic next-nearest-neighbor coupling $J_2$ [illustrated in Fig.~\ref{fig:intro}(a)]. For all $N\geq 5$,
this model harbors a VBS phase for $J_2/J_1=0$ and a N\'eel phase for $J_2/J_1=\infty$. By detailed unbiased QMC 
studies for $5\leq N\leq 12$, we find that that the two phases are separated by a single phase transition, 
with no signs of discontinuities even on the largest systems sizes studied ($L\times L$ spins with $L$ up to $128$). 
Most remarkably, the anomalous dimensions of the N\'eel and VBS correlation functions of the model for large-$N$ 
shows quantitative agreement with the analytically known \cite{murthy1990:mono,metlitski2008:mono,kaul2008:u1}
scaling dimensions from the $1/N$ expansions of the non-compact CP$^{N-1}$ model. 

{\em The $J_1$-$J_2$ model.}---Our SU($N$) symmetric model is defined with a local Hilbert space of $N$ states on 
each site of the square lattice illustrated in Fig.~\ref{fig:intro}(a). We adopt the representation used previously
in both analytic \cite{read1989:vbs} and numerical \cite{harada2003:sun,beach2009:sun} works on bipartite lattices,
where the sublattice-A states transform under rotations with the fundamental representation of SU($N$), and the B sublattice 
states transform with the conjugate of this representation; 
$|\alpha\rangle_A\rightarrow U^{~}_{\alpha\beta}|\beta\rangle_A$,
$|\alpha\rangle_B \rightarrow U^*_{\alpha\beta}|\beta\rangle_B$. 
The state $\sum_\alpha |\alpha\rangle_A|\alpha\rangle_B$ is, thus, an SU($N$) singlet. $P_{ij}$ is defined to be the projector 
onto this singlet between two sites $i$ and $j$ on {\em different} sublattices, i.e., $H_{ij}=-P_{ij}/N$ is the SU($N$) generalization 
of the familiar Heisenberg antiferromagnetic exchange (up to a constant). Another simple SU($N$) invariant interaction is the 
permutation operator between two sites on the {\em same} sublattice, $\Pi_{ij}|\alpha \beta\rangle=|\beta\alpha\rangle$, 
so that  $H_{ij}=-\Pi_{ij}/N$ is the generalization of the the familiar ferromagnetic Heisenberg interaction. The Hamiltonian 
we study here is given by
\begin{equation}
\label{eq:j1j2N}
H= -\frac{J_1}{N}\sum_{\langle ij \rangle}P_{ij}
-\frac{J_2}{N}\sum_{\langle\langle ij \rangle\rangle}\Pi_{ij},
\end{equation}
where $\langle ij \rangle$ and $\langle\langle ij \rangle\rangle$ denote first
(A-B) and second (A-A and B-B) neighbor sites, respectively. 

With $J_2=0$ it is now well known that the $J_1$ model 
is N\'eel ordered for $N=2,3,4$ and develops VBS order for $N\geq 5$ \cite{harada2003:sun,beach2009:sun}. On the other hand, 
with $J_1=0$ each sublattice forms a trivial ferromagnet. A small $J_1\ll J_2$ will clearly lock the individual sublattice 
magnetizations into a collective N\'eel ordered state. Thus, for each $N>5$ there must be an intermediate value of 
$g\equiv J_2/J_1$ at which there is a quantum transition between these two phases (as we do not expect any other
intervening phase). 

{\em QMC simulations.---}All off-diagonal matrix elements in Eq.~(\ref{eq:j1j2N}) are explicitly negative 
and, hence, the model is free of QMC sign problems [and it also satisfies Marshall's sign criterion, ensuring 
an SU($N$) singlet ground state]. To obtain exact (within statistical errors) numerical 
results for its properties on large $L\times L$ lattices, we use the stochastic series expansion QMC method with 
global loop updates \cite{sandvik1999:sse,evertz2003:loop,sandvik2010:vietri}. Throughout this work, we set $J_1=1$ and the inverse 
temperature $\beta=L/J_1$ (reflecting the expected \cite{senthil2004:science} dynamic exponent $z=1$). 

We characterize the N\'eel phase as one with a finite spin stiffness $\rho_s$ (measured by the fluctuations of the winding
number $W$ of world lines; $\beta\rho_s = \langle W^2\rangle$ \cite{{pollock1987:winding},sandvik2010:vietri}).
In the magnetic phase, the static ($\omega=0$) N\'eel order-parameter susceptibility $\chi_N$ diverges with the 
``quantum volume'' of the system according to $\chi_N\sim \beta L^2$. We define the SU($N$) VBS correlation function 
using the operator $P$ defined above in Eq.~(\ref{eq:j1j2N}); 
$C_V({\bf r},\tau)=\langle P_{\bf 0,0+x}(0) P_{\bf r,r+x}(\tau) \rangle-\langle P_{\bf 0,0+x}(0) \rangle^2$. When Fourier 
transformed at $\omega=0,{\bf q}=(\pi,0)$ it gives $\chi_V$. This quantity can be used to test for VBS order since it diverges 
in the VBS phase as $\chi_V\sim \beta L^2$. We also use the standard definition of the correlation length of 
the VBS order $\xi_V$ as the square root of the second moment of the spatial correlation function $C_V$. Using these quantities 
we tested for long-range N\'eel and VBS order as the ratio $g=J_2/J_1$ is varied for each $N$ and arrived at the phase diagram 
shown in Fig.~\ref{fig:intro}(b). We elaborate on the quantitative analysis below.

\begin{figure}[t]
\includegraphics[width=8cm, trim = 20 60 0 0]{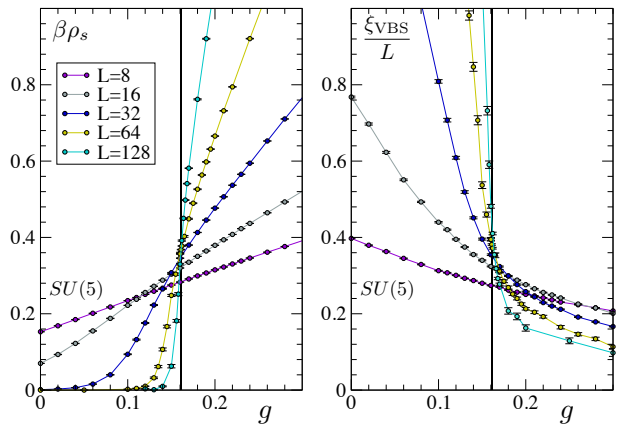}
\caption{\label{fig:su5cr} (Color online) Curve crossings used to locate the critical point for 
  magnetic [VBS] order in the SU(5) $J_1$-$J_2$ model are shown in the left [right] panel. The quantity 
  $\beta \rho_s$ [$\xi_{\rm VBS}/L$] diverges in the magnetic [VBS] phase and goes to zero in 
  the non-magnetic [non-VBS] phase when $\beta = L/J_1$. At a point where magnetic [VBS] 
  fluctuations are critical, $\beta \rho_s$ [$\xi_{\rm VBS}/L$] becomes $L$-independent. These 
  properties result in crossings of curves for different $L$ at the critical point. The width of
  the vertical line shows the range of estimates of the common N\'eel-VBS critical point; 
  $g_{\rm c} =1.615(10)$. Fig.~\ref{fig:cross} shows the analysis of the crossing points 
  giving this result.}
\vskip-4mm
\end{figure}

{\em Nature of the phase transition.}---Fig.~\ref{fig:su5cr} shows QMC results for $\beta \rho_s$ and 
$\xi_{\rm VBS}/L$ as functions of the coupling ratio $g$ for the SU($5$) model on lattices of size
$L=8,16,32,64$, and $128$. The quantum-critical point for the magnetic and VBS orders can be located by analysing 
crossing points versus $g$ in $\beta \rho_s$ and $\xi_{\rm VBS}/L$, respectively, computed on two different system 
sizes. As is clearly evidenced directly from this raw data, there are crossing points within a narrow window of 
$g$ for both N\'eel and VBS orders, and these crossing points drift toward a common value $g_c$ with increasing $L$.

\begin{figure}[t]
\includegraphics[width=7.5cm, trim = 20 17 0 0]{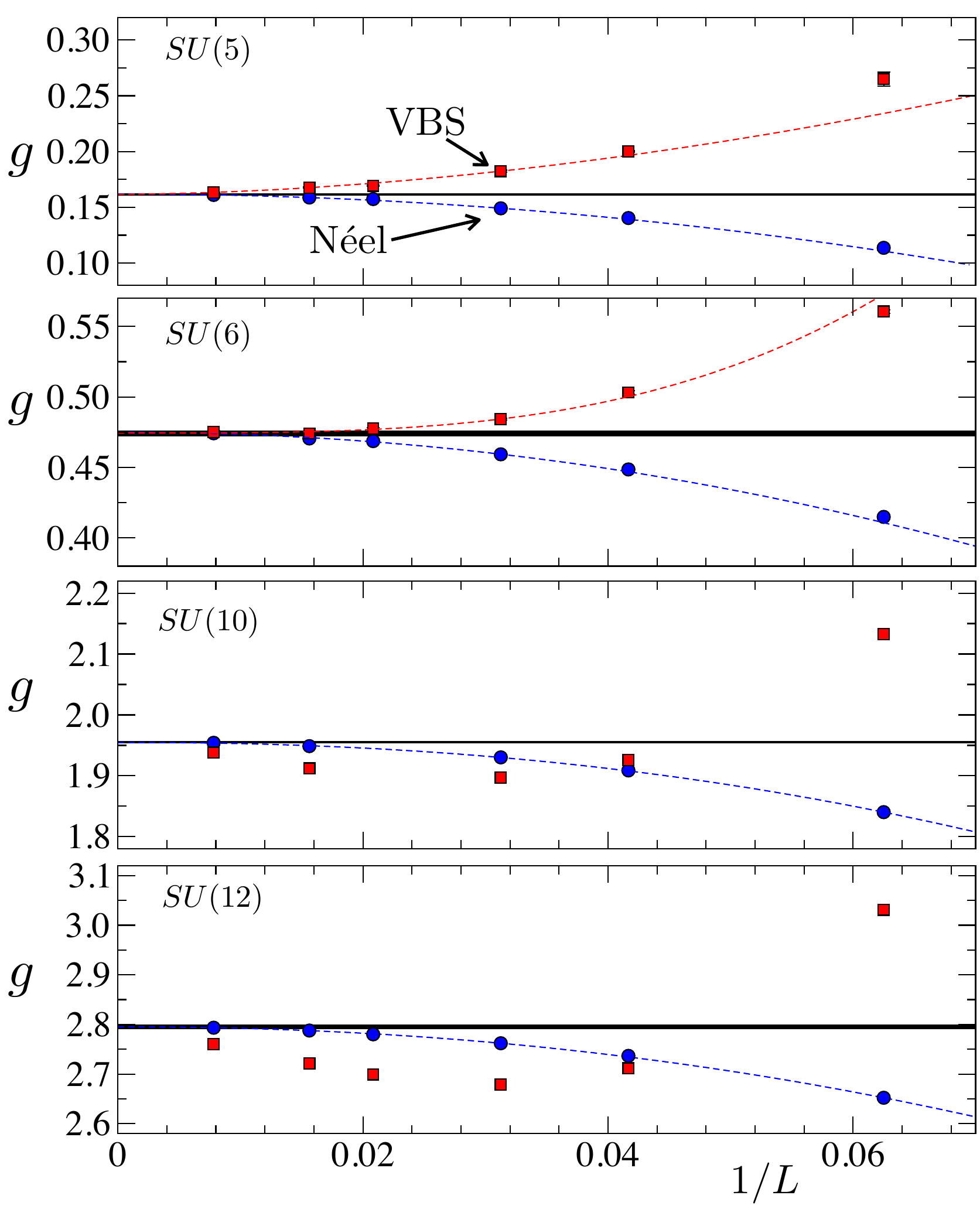}
\caption{\label{fig:cross} (Color online) Convergence to a common critical point of the finite-size estimates for
  the N\'eel and VBS phase transitions for SU($5$), SU($6$), SU($10$), and SU($12$). The crossing points of $L$
  and $L/2$ curves for the same quantities as in Fig.~\ref{fig:su5cr} are shown for N\'eel (blue circles) and VBS (red squares) 
  order for $L\leq 128$. For all four cases, the data are consistent, within error bars, with both order parameters becoming 
  critical at the same point when $L\to\infty$. Numerical extrapolations (when reliable) are shown by dashed lines. For 
  SU($10$) and SU($12$) the VBS correlation function has significant subleading corrections that can be ignored only for 
   the largest sizes (see \cite{corrections}). The grey lines show our
   estimates of the critical points (with their widths 
   corresponding to the error bars). }
\vskip-4mm
\end{figure}

In Fig.~\ref{fig:cross} we have plotted the crossing points between $L$ and $L/2$ curves of $\beta \rho_s$ 
and $\xi_{\rm VBS}/L$ for SU($N$) systems with $N=5$, $6$, $10$, and $12$. Numerical extrapolations 
of the crossing data for both N\'eel and VBS orders in the SU($5$) and SU($6$) cases  (top two panels)
provide compelling evidence that in the thermodynamic limit the crossing points for both order-parameters 
approach a common critical point. For SU($10$) and SU($12$) (bottom two panels) there is non-monotonicity 
in the VBS crossing points, making reliable numerical extrapolations difficult. The source of this behavior 
is subleading corrections in the VBS correlation function that increasingly dominate the rapidly decaying leading 
power-law behavior as $N$ increases (see note \cite{corrections}). However, even for the SU($10$) and SU($12$) cases, 
for the largest system sizes the leading behavior still dominates the subleading corrections, and the VBS crossings 
are consistent with the same critical point as obtained from the better-behaved $\beta \rho_s$ crossing in the 
$L\rightarrow \infty$ limit. There is, thus, good reason to believe that both orders go critical at the same point 
for all $N$. We have found no evidence for hysteresis or double peaked histograms that would be expected for a
first-order transition.

Since a reliable numerical extrapolation of the $\beta \rho_s$ crossing is possible for each $N$, we use this 
quantity to extract the quantum-critical points. Note that possible weak corrections to standard scaling 
behavior in quantities whose scaling from depends only on the dynamic exponent, such as $\rho_s$, which have been discussed for the J-Q model \cite{sandvik2010:logs,kaul2011:su34,banerjee2010:log}, 
would not affect the above analysis of {\em crossing points} versus $g$ (see
Appendix D in \cite{kaul2011:su34}). Corrections to scaling of the quantity $\beta \rho_s$ 
itself will be studied in detail elsewhere. 

\begin{figure}
\includegraphics[width=7.25cm, trim = 20 32 0 0 ]{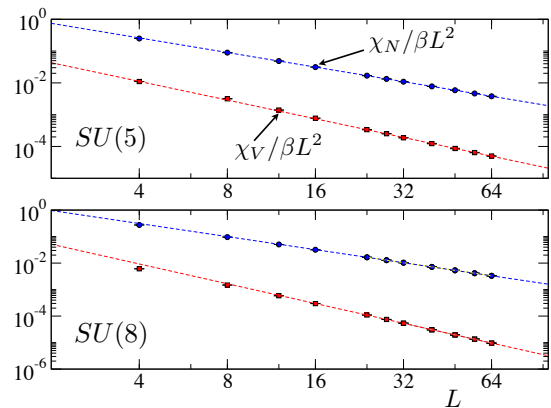}
\caption{\label{fig:powerNV} (Color online) Examples, for the SU($5$) and SU($8$) models, of power-law scaling of the N\'eel 
  and VBS order parameters close to the quantum-critical point. According to Eq.~(\ref{eq:expdef}) and standard finite-size
  scaling arguments, $\chi_N/\beta L^2 \sim 1/L^{1+\eta_N}$  and $\chi_V/\beta L^2 \sim 1/L^{1+\eta_V}$. The linear regression 
  fits (dashed lines) allow us to extract $\eta_V$ and $\eta_N$, which are shown in Fig.~\ref{fig:exp} as a function of $N$.}
\vskip-2mm
\end{figure}

At a common quantum-critical point, it is expected that the otherwise independent N\'eel and VBS order
parameters will {\em both} have correlation functions that decay as power laws in
space (${\bf r}$) and imaginary time ($\tau$);
\begin{equation}
\label{eq:expdef}
C_{N,V} ({\bf r},\tau) \sim  {(\bf r}^2+c^2\tau^2)^{-(1+\eta_{V,N})/2}.
\end{equation}
The exponents $\eta_N$ and $\eta_V$, the so-called ``anomalous dimension'' of the N\'eel and VBS fields, are universal 
numbers according to standard theory of critical phenomena. Universality implies that they are independent of details 
of the microscopic interactions of the model from which the correlation functions are extracted, but they do depend on 
the symmetry of the model, {\em i.e.}, in our case they should only depend on $N$ of the SU($N$) symmetry. To estimate 
these exponents with QMC simulations, we have studied the $N=5,6,8,10$, and $12$ models at values of the coupling ratio 
$g$ within the estimated critical points from the analysis shown in Fig.~\ref{fig:cross}. We extract the exponents 
from the size dependence of the correlation functions, as illustrated for two cases in Fig.~\ref{fig:powerNV}.

Here we note again that the scaling behavior in the J-Q model of quantities depending only on the dynamic exponent
show  weak corrections to scaling 
\cite{sandvik2010:logs,banerjee2010:log,kaul2011:su34}. If such corrections are present also in the correlation functions 
analyzed here, they would somewhat affect the values of the exponents,
{\em e.g.}, a multiplicative log is hard to distinguish 
from a small change in an exponent (see \cite{sandvik2010:vietri} for an example in one-dimension).

\begin{figure}[t]

\includegraphics[width=7.5cm,  trim = 20 35 0 0 ]{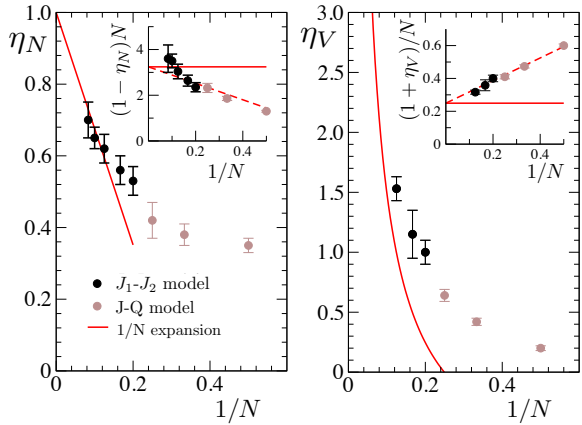}
\caption{\label{fig:exp} (Color online) Anomalous dimensions of the N\'eel (left) and VBS (right) fields extracted from 
  the critical scaling analysis. The main panels show $\eta_N$ and $\eta_V$ versus $1/N$. For $N=2,3$ and $4$, the data are 
  for the J-Q model \cite{lou2009:sun}, and the results for $N>4$ are for our $J_1$-$J_2$ model. The analytic results 
  from the $1/N$ expansion of the CP$^{N-1}$ field theory are shown as thick red lines. The left and right insets 
  show $N(1-\eta_N)$ and $(1+\eta_V)/N$, respectively. These quantities must be finite in the  $N\rightarrow \infty$ 
  limit according to the DQC theory and should be given by Eq.~(\ref{eq:oneonN}) (solid straight lines in the insets). 
  The next corrections to the exponents have not been computed analytically yet, but we can estimate them approximately 
  as $1+\eta_V = 0.2492 N + 0.68(4)$, $\eta_N = 1+32/(\pi^2 N)-3.6(5)/N^2$ (shown as dashed lines).}
\vskip-2mm
\end{figure}

{\em Comparison with large-$N$ results.}---The extracted exponents $\eta_N$ and $\eta_V$ are shown versus $1/N$ in the main panels 
of Fig.~\ref{fig:exp}. For SU($10$) and SU($12$),  $\eta_V$ becomes too large to extract reliably (see note \cite{corrections}). 
Our main objective is to compare the results with those of the CP$^{N-1}$ universality predicted by the DQC scenario 
\cite{senthil2004:science}. Analytic large-$N$ results currently available for these indices are
\begin{equation}
\label{eq:oneonN}
\eta_N = 1 - 32/(\pi^2N),~~~~1+\eta_V = 2 \delta_1 N,
\end{equation}
where $\delta_1\approx 0.1246$. It is interesting to note that the calculations underlying these two results are entirely 
different. While $\eta_N$ was obtained from an $1/N$ expansion of the N\'eel order parameter expressed in terms of the 
CP$^{N-1}$ fields~\cite{kaul2008:u1}, $\eta_V$ was computed \cite{murthy1990:mono,metlitski2008:mono} by exploiting a 
non-trivial relation, predating the DQC theory, between monopoles in the field theory and the VBS order on the lattice
\cite{read1989:vbs}. 

In Fig.~\ref{fig:exp} we show the results of the $1/N$ expansion, Eq.~(\ref{eq:oneonN}), as continuous curves. 
In the insets we plot the same data in such a way that in the $1/N\rightarrow 0$ limit we can do a 
more direct comparison with the two irrational numbers $32/\pi^2$ and $2\delta_1$ that are predicted based on the CP$^{N-1}$ 
theory. We have also fitted these data to straight lines, which give numerical predictions of the next-higher corrections 
to the large-$N$ forms in Eq.~(\ref{eq:oneonN}).  

The most important features in Fig.~\ref{fig:exp} that lend support to the DQC scenario are: {\em (1)} The 
exponents $\eta_N$ and $\eta_V$ for the $N>4$ $J_1$-$J_2$ models are consistent in trend with previous estimates for 
$N=2,3,4$ based on the entirely different J-Q model~\cite{lou2009:sun}. This is line with the concept of universality 
between different microscopic models, characteristic of a continuous transition. {\em (2)} The N\'eel-order exponent $\eta_N$ 
connects to the leading $1/N$ behavior, approaching $1$ as $N\rightarrow\infty$. This large value (in contrast to the
normal mean-field value $\eta=0$ and typically very small values at conventional critical points) was one of the important 
early ``smoking gun'' predictions of the DQC scenario~\cite{senthil2004:science}. {\em (3)} The leading $1/N$ correction 
to $\eta_N$ is of the correct sign and within a few percent in
magnitude of the analytic result $32/\pi^2$ for the CP$^{N-1}$ 
theory \cite{kaul2008:u1}. {\em (4)} $\eta_V$ increases rapidly with 
$N$ and has a trend that is fully consistent with the very non-trivial large-$N$ result
\cite{murthy1990:mono,metlitski2008:mono,kaul2008:u1}.  

{\em Conclusions.}---We have provided an SU($N$) symmetric sign problem free model that allows, for the first time, 
unbiased studies of the N\'eel-VBS transition for arbitrary $N>5$ on large lattices. Our model generalizes to SU($N$) the important
$J_1$-$J_2$ model for SU($2$) spins, which has a long history of studies by exact diagonalization and various approximate methods
(normally for antiferromagnetic $J_2$, in which case it has a N\'eel--VBS transition also for $N=2$, which, however, is difficult
to study reliably due to QMC sign problems) \cite{j1j2refs}. Our model opens new avenues for detailed connections between large-$N$ 
theory and unbiased numerical simulations.

In this initial study, we found direct N\'eel-VBS transitions for $5\leq N \leq 12$, with no signs of discontinuities up to large 
system sizes ($L\leq 128$). When analyzed as continuous quantum critical points, we fond remarkable agreement for universal exponents 
with the $1/N$ expansion of the CP$^{N-1}$ field theory. This quantitative comparison lends very strong support to the DQC theory 
\cite{senthil2004:science}. 

{\em Acknowledgments.}---We gratefully acknowledge inspiring discussions with Michael Levin. Partial financial support was 
received through NSF Grants No.~DMR-1056536 (RKK) and DMR-1104708 (AWS). The numerical simulations were carried out on the 
DLX cluster at the University of Kentucky.  

\null


\begin{thebibliography}{00}
\bibitem{corrections}
As is apparent from Fig.~\ref{fig:exp} and Eqs.~(\ref{eq:expdef},\ref{eq:oneonN}), the large corrections to scaling in the VBS correlation 
function $C_V(r)$ are due to the fact that $\eta_V$ itself becomes large as $N$ is increased, resulting in a $C_V(r)$ whose leading term decays 
too quickly to separate it from subleading pieces on moderate-size lattices and with typical QMC errorbars.
\end{thebibliography}
\end{document}